\documentclass[nonacm]{acmart}
\usepackage{multirow}
\usepackage{listings}
\usepackage{ulem}

\setcopyright{rightsretained}
\acmPrice{}
\acmDOI{}
\acmYear{2020}
\copyrightyear{2020}
\acmJournal{PACMPL}
\acmVolume{1}
\acmNumber{SYNT}
\acmArticle{2}
\acmMonth{7}

\acmConference[SYNT '20]{SYNT '20: 9th Workshop on Synthesis}{July 19, 2020}{Los Angeles, CA, USA}
\acmISBN{}

\lstdefinelanguage{alg}{
  sensitive = true,
  keywords={algorithm, input, output, return, if, else, for, while},
  numberstyle=\footnotesize,
  numbersep=9pt,
  showstringspaces=false,
  breaklines=true,
  comment=[l]{//},
  morecomment=[s]{/*}{*/},
}

\lstdefinestyle{grammar}{
  belowcaptionskip=1\baselineskip,
  basicstyle=\rmfamily\mdseries\footnotesize,
  breaklines=true,
  language=alg,
  xleftmargin=\parindent,
  showstringspaces=false,
  mathescape=true,
  numberstyle=\tiny,
  keywordstyle=\bfseries
}

\begin{document}




\title{Gradient Descent over Metagrammars for Syntax-Guided Synthesis}

\author{Nicolas Chan}
\email{nicolaschan@berkeley.edu}
\affiliation{%
  \institution{UC Berkeley}
}

\author{Elizabeth Polgreen}
\email{epolgreen@berkeley.edu}
\affiliation{%
  \institution{UC Berkeley}
}

\author{Sanjit A. Seshia}
\email{sseshia@berkeley.edu}
\affiliation{%
  \institution{UC Berkeley}
}

%
\begin{abstract}
Syntax-guided synthesis restricts the search space for program synthesis through
the use of a grammar imposing a syntactic restriction.
In practice, provision of such a grammar is
often left to the user to do manually, though in the absence of such a grammar, state-of-the-art solvers will
provide their own default grammar, which is dependent on the signature of the target program to be sythesized.  
In this work, we speculate this default grammar could be improved upon substantially. 
We build sets of rules, or metagrammars, for constructing grammars, and perform a gradient descent over these metagrammars aiming to find a metagrammar which solves more benchmarks and on average faster.
We show the resulting metagrammar enables CVC4 to solve $26$\% more benchmarks than the default grammar within a $300$s time-out, and that metagrammars learnt from tens of benchmarks generalize to performance on 100s of benchmarks. 
\end{abstract}

\maketitle

\section{Introduction}
 A major theme in the advances in program synthesis over the past 15 years has been the use of syntactic restrictions on the space of programs being searched. One prominent approach is that of Syntax-Guided Synthesis~\cite{DBLP:series/natosec/AlurBDF0JKMMRSSSSTU15}, which
augments the program synthesis problem with a syntactic template or grammar from which the program is
to be constructed. This template can be used to guide the synthesis algorithm and to reduce the search
space of possible solutions. 

Provision of this grammar, however, typically requires a skilled domain expert.
If the
grammar is too small, it may not be expressive enough to contain a solution to the synthesis task, 
or any
solution might be intractably large. For instance, 
constant literals may need to be constructed as the sum of smaller constant literals e.g., $1's$.
%
%
If the grammar is large, the space of potential solutions is large
and so the run-time of the synthesis algorithm will be longer.
Furthermore, a grammar may contain some operators that are more expensive for a synthesis algorithm to reason about than others.
Consequently, choosing a good grammar is a key part in the success of SyGuS. 

The best particular grammar for a synthesis problem will depend on the function
to be synthesized, for example, it will depend on the number and type of arguments of the function, 
the
return type of the function, or any number of other particularities about the
synthesis problem. 
So to construct a good grammar in general, we need
a set of rules that output a good particular grammar for a particular
synthesis problem.

In this work, we call this set of rules a \textit{metagrammar}. The metagrammar may include
rules such as adding the function arguments to the grammar, adding various
functions (such as \texttt{bvadd}, \texttt{bvor}, etc. for bitvector problems), or
adding helper functions and constants present in the constraints or other parts
of the problem statement.

We design a scoring system to reward metagrammars solving benchmarks quickly
and penalize those failing to solve benchmarks, and use a simple gradient descent over these metagrammars. We evaluate the resulting metagrammar on a set of bitvector benchmarks from the SyGuS competition, using CVC4~\cite{DBLP:journals/corr/abs-1806-08775} as the SyGuS solver, and find metagrammars learnt from a randomly selected training set of tens of benchmarks,
are able to solve $26$\% more benchmarks than the default grammar on the remaining benchmarks within a 300$s$ time-out.

\subsection{Related work}
The balance between expressivity of a grammar and the speed of synthesis is discussed by Gulwani in the context of data-wrangling~\cite{DBLP:series/natosec/Gulwani16}, and further by Padhi et al, who hypothesise that the performance of SyGuS tools degrades as the expressivity of the grammar increases, and propose a solution based on interleaving multiple grammars~\cite{DBLP:conf/cav/PadhiMN019}.
Morton et al.~\cite{DBLP:journals/corr/abs-2002-02884} use machine learning to filter the 
grammars used by SyGuS.
This work assumes a grammar is given, and optimizes that grammar, whereas we assume no grammar is provided
and improve upon the default grammar. 
A key difference is that Morton et al. use a neural network to perform the filtering and so need a large amount of training data. The authors test their techniques on programming-by-example benchmarks (benchmarks where the specification is a set of input-output examples which the desired program satisfies), where automatically generating new benchmarks is straightforward. A key benefit of our approach is that it requires very small amounts of training data, which is difficult to obtain for program synthesis for logical specifications (where the desired program must satisfy a set of logical constraints). It is not straight forward to convert a logical specification into a programming-by-example specification~\cite{DBLP:journals/corr/abs-2001-09245,DBLP:journals/corr/abs-1710-01291}.
Machine learning based program synthesis techniques~\cite{DBLP:journals/corr/abs-1805-04276,DBLP:conf/icml/DevlinUBSMK17}, which often do look at identifying important pieces of a grammar~\cite{DBLP:conf/iclr/BalogGBNT17,DBLP:conf/iclr/KalyanMPBJG18}, are usually applied to PBE problems, in part due to this challenge. A key advantage of our simple gradient descent based technique is that we do not need large amounts of training data in order to achieve performance gains. 

Research into techniques used for selecting grammars for SyGuS and research into techniques used for enumerating through the grammars~\cite{DBLP:conf/cav/ReynoldsBNBT19,DBLP:conf/tacas/AlurRU17} tackle similar problems, and good enumeration strategy should mitigate the effects of a grammar specifying a large search space. An interesting example is EUPHONY~\cite{DBLP:conf/pldi/LeeHAN18}, which uses learns a probabilistic model at run time that biases the enumeration. However, there are advantages to having an algorithm for selecting grammars as well as an effective enumeration algorithm. It provides flexibility as the algorithm can be treated as a bolt-on for any SyGuS solver.  It is also likely that the best metagrammar will be be dependent on the set of benchmarks you wish to solve. We can evision a scenario where a user has bodies of synthesis problems from different specific applications and uses our technique to generate a metagrammar for use with each specific benchmark set. Our framework could also allow a user to specify rules as part of the metagrammar which extract features the user knows to be important from the benchmark.


\section{Background}
\subsection{Syntax-Guided Synthesis}
A Syntax-Guided Synthesis~\cite{DBLP:series/natosec/AlurBDF0JKMMRSSSSTU15} (SyGuS) problem consists of two constraints, a semantic constraint
given as a formula $\phi$, and a context-free grammar, often referred to as the syntactic template. The computational problem is then to find a function $f$, built from the context-free grammar, such that 
$$\forall x\,\,\phi(f,x)$$ is valid, where $x$ is the set of all possible inputs to $f$. 
\subsection{Metagrammars}
A metagrammar is a set of rules which determine what to add to the
grammar for a particular synthesis problem.
A rule is a mapping from a particular synthesis specification $\phi$ to a set of
terminal symbols \(S\), where the specification includes the signature for the function $f$ to be synthesized. For example, a rule specifying that the arguments of the
function are included in the grammar would inspect $\phi$, extract the function
arguments, and then output \(S\) consisting of the arguments as terminals. Then each
terminal in \(S\) can be included in the definition of the appropriate
non-terminal to construct the final grammar for the particular 
synthesis problem (in union with the other rules).

\section{Methodology}
\subsection{Building Metagrammars}
\label{sec:metagrammars}
We begin by identifyinga default grammar which includes a non-terminal for each data type in the function signature plus
booleans and the
following rules for populating these non-terminals:
\begin{itemize}
\item Add argument variables of the same data type
\item Add constants $1$ and $0$ for each data type
\item Add functions (operators) for each data type
\item Add predicates for each relevant data type to the \texttt{Bool} type (for use in the
condition of if-then-else)
\end{itemize}
%
We identify a set of rules for producing the default grammar, where each
rule adds a set of constants and/or operators to the grammar.
This set of rules forms the default metagrammar which we can use to generate a grammar for a given synthesis problem. 
We use this basic set of rules to evaluate our framework but more complex rules could be explored in future work.
We could, for instance, introduce rules which extend the grammar, for example rules which add constants and  
helper functions found in the synthesis specification file.



\begin{table*}
\begin{center}
\begin{tabular}{l | l | c  c | c c c | c c c}
&  & \multicolumn{2}{c}{Default metagrammar} &\multicolumn{3}{c}{Reduced metagrammar}  &\multicolumn{3}{c}{Enhanced metagrammar}\\\hline
 Set               & total &solved&avg. time&solved&unique&avg. time&solved&unique&avg. time\\ \hline
 Non-SI bitvector  & 421   & 276  &  18.1$s$&  271 &  12  &  13.3$s$& 282  &  21  & 12.5$s$\\
 Non-SI (orig LIA) & 429   & 165  &  14.9$s$&  181 &  19  &  13.6$s$& 165  &   0  & 15.8$s$\\
 PBE bitvector     & 705   & 244  &  33.3$s$&  379 &  146 &  40.8$s$& 556  &  331 & 1.0$s$\\\hline 
 Total             & 1555  & 685  &  22.7$s$&  831 &  177 &  25.9$s$& 1003 &  352 & 6.6$s$ \\
\end{tabular}
\end{center}
\caption{The number of benchmarks solved and average solving time in seconds for the default, reduced and enhanced metagrammars. The unique columns show the number of benchmarks solved, that were not solved by the default grammar.
We discount the performance of the enhanced metagrammar on the PBE benchmarks because adding in constant literals to a grammar allows the solver to produce large
look-up tables as solution. 
\label{tab:results}}
\end{table*}


\subsection{Metagrammar Search}
\label{sec:org884fb50}
We score each metagrammar according to its performance on a
set of training benchmarks. Where the benchmarks are easily divisible into 
representative categories, for instance invariant generation, we stratify the training data across these sets in order to attempt to obtain
representative training data.
Based on these scores we perform a search by expanding the
neighbors of the best performing metagrammar.

As described, a metagrammar \(M\) is a set of rules \(M=\{r_{1}, r_{2}, \dots, r_{n}\}\). A
\textit{smaller neighbor of \(M\)} is a metagrammar \(N_{i}\) which is missing one of the rules.
That is, the set of smaller neighbors of \(M\) is \(\{ N_{i} = M\setminus \{r_{i}\} : i
= 1,2,\dots,n \}\).

The search begins with the neighbors of the default metagrammar. Then the neighbors of the best neighbor are expanded, and so on, until we have
exhausted all of the rules.

\subsubsection{Scoring}
\label{sec:org96db1da}
We designed a scoring function that aims to reward metagrammars
solving a benchmark significantly faster than neighbor metagrammars, and heavily penalize metagrammars
failing to solve a benchmark. We base this on the normalized runtime against the neighbors. 
%
The score $S_B$ for a
particular metagrammar $M$ on a benchmark $B$, 
in the context of the results of the $n$
neighboring metagrammars, for a single benchmark is determined by the difference of $M$'s runtime against the average of its neighbors normalized by the standard deviation:
$$S_B = \begin{cases}
10 & \text{benchmark unsolved,} \\
\frac{r_{M} - \frac{1}{n}\sum_{0 \leq j \leq n}r_{N_j}}{\sigma} & \text{benchmark solved} 
\end{cases}$$
where $r$ is the runtime of the metagrammar in seconds and $\sigma$ is the standard deviation
of the runtime of all the metagrammars on that benchmark. A lower score is better. 
%
%
The specific scoring system is a heuristic and may be improved upon or tailored for specific use cases, for instance one could incorporate a metric that rewards finding shorter solutions, or finding solutions that contain certain operators. 

\section{Evaluation}
\label{sec:org512a52f}
We use CVC4 1.7 (\texttt{84da9c0b}) with the
\texttt{--cegqi-si=none} option,
and remove single invocation benchmarks from our benchmark set since these do not benefit 
from a grammar being provided.
We use a 300 second timeout, with 20 benchmarks running in
parallel on each Intel Xeon E5-2670 v2 CPU. We use a set of $1699$ 
bitvector benchmarks taken from the SyGuS competition~\cite{sygus-comp},
grouped into $3$ sets, shown in Table~\ref{tab:results}: non single-invocation (non-SI) bitvector benchmarks;
 programming-by-example (PBE)
bitvector benchmarks; and non-SI Linear Integer Arithmetic benchmarks
that we translate into bitvectors (this includes invariant generation benchmarks).

We extract the default grammar used by CVC4\footnote{
Obtained from CVC4 version 1.7 using the \texttt{-{}-trace=sygus-grammar-def} option.}
and identify a default metagrammar capable of producing this grammar. 
We run the gradient descent process starting with the default metagrammar.
We train on a stratified 
sample of benchmarks, taking 48 from each of the 3 sets shown in Table~\ref{tab:results}. We stratify the training data across these 3 sets in order to attempt to obtain
representative training data. These categories are very broad though and better results may be obtained by categorising the benchmarks into more precise categories, or by training a metagrammar for each set,
We obtain a metagrammar which we call the \textit{reduced metagrammar}, and which produces the following
grammar for a function with bitvector arguments A1 and A2:
\begin{lstlisting}[style=grammar,language=alg] 
BV ::= A1|A2|1|(ite Bool BV BV)|(bvnot BV BV)|
       (bvor BV BV)|(+ BV BV)|(bvlshr BV BV)| 
       (bvshl BV BV) 
predicate ::= (= BV BV)|(bvult BV BV)
Bool ::= true|(not Bool)|(and Bool Bool)| 
         (or Bool Bool)|predicate
\end{lstlisting}
We also take the default metagrammar and manually add a rule adding 
constant literals found in the benchmark,
and we call the resulting metagrammar the \textit{enhanced metagrammar}.  

We evaluate all metagrammars over the remaining set of $1555$ benchmarks. 
The reduced metagrammar 
solved $177$ benchmarks that the default did not
solve, while failing to solve 31 benchmarks the default metagrammar did solve.
It is $22$\% faster on the first two categories of benchmarks, but slower on the PBE category.
The enhanced metagrammar solves $21$ benchmarks that the default did not solve, taken
from the bitvector category, but does is on average slower and solves no new benchmarks in the
category of benchmarks translated from LIA (which includes all benchmarks from the invariant synthesis
category). We hypothesize that this is because, whilst adding constant literals to a grammar is helpful,
it is
not sufficient to do so without also reducing the size of the grammar when the benchmarks are more complex.
 This suggests it would be beneficial to explore using gradient descent over the
space of metagrammars that include rules that extend beyond the default metagrammar,
with the aim of combining the success of the reduced and enhanced metagrammars.
We discount the results for the enhanced metagrammar on the PBE category since adding in constants
makes it possible for the solver to effectively produce lookup table solutions to these benchmarks.
The differences in results across the categories suggest it would be worth further 
exploring stratified sampling for training sets,
learning metagrammars based on training sets taken solely from the category 
that the metagrammar is going to be tested on, or metagrammars containing rules that attempt
to classify benchmarks into such categories. 

It is worth noting that in some cases the learnt metagrammar results in faster solving but potentially longer solutions. This would not necessarily be a desirable outcome, but it this behaviour is expected given that our scoring function only rewards speed of solving and number of benchmarks solved. A scoring function that places some weighting on ``quality'' of solutions would be interesting to explore in future work. It is also worth noting that a different training set will result in a different metagrammar; we report results from a randomly selected training set. A fuller evaluation is in order to determine the optimum size of the training set in order to consistently produce these improvements, but we note that we trained two further metagrammars which both improved over the default (solving 744 benchmarks in an average of $22.0$s, and solving 814 benchmarks in an average of $27.7$s, with significant improvements in the Non-SI bitvector and Non-SI LIA translated categories respectively).

\section{Conclusions}
We have presented a framework that uses a gradient descent based search 
to find a metagrammar that improves solver performance over 
a set of SyGuS benchmarks. 
We have evaluated one 
specific instance of this framework, 
using bitvector benchmarks and the specific set of metagrammar
rules and the scoring function described in this paper. Despite its simplicity, and the 
minimal amount of training data benchmarks required, we 
found this instantiation of our method solved $25\%$ more benchmarks 
than the default metagrammar within
the timeout. 
In future work we plan to explore more advanced gradient descent algorithms, 
as well as incorporating more advanced rules in the metagrammars and exploring 
the relationships between metagrammars and different enumerative synthesis algorithms. 
We believe that techniques like this could have significant impact in application domains for synthesis
such as synthesising invariants for a specific type of software,
where benchmarks are available but not in the abundance required for neural network based techniques.

\paragraph{Acknowledgements}
This work was supported in part by NSF grants 1739816 and 1837132, a gift from Intel under the SCAP program, and the iCyPhy center.

\bibliography{paper}{}
\bibliographystyle{plain}
\end{document}